\documentclass{elsart}
\usepackage{graphicx}
\usepackage{amssymb}
\def\beq{\begin{eqnarray}}
\def\eeq{\end{eqnarray}}

\begin{document}

\begin{frontmatter}

\title{Development of accurate solutions for a classical oscillator}
\author{Paolo Amore}
\ead{paolo@ucol.mx}
\address{Facultad de Ciencias, Universidad de Colima,\\
Bernal D\'{i}az del Castillo 340, Colima, Colima,\\
Mexico.}
\author{Nestor E. Sanchez}
\address{Department of Mechanical Engineering, The University of Texas at San Antonio,\\
San Antonio, TX 78249}

\begin{abstract}    
We present a method to obtain arbitrarily accurate solutions for conservative
classical oscillators. The method that we propose here works both for 
small and large nonlinearities and provides simple analytical approximations.
A comparison with the standard Lindstedt-Poincar\'e method is presented, from which 
the advantages of our method are clear. 
\end{abstract}

\end{frontmatter}


\section{Introduction}

In this paper we consider the problem of calculating the periodic solutions to
the differential equation 
\begin{equation}
m \frac{d^2x}{dt^2}+f(x)=0 \ .
\label{eqn1}
\end{equation}

Eq.(\ref{eqn1}) is the Newton equation for a particle of mass $m$ moving under the
action of a force $f(x)$. We assume that the system is conservative and therefore we
introduce the potential $V(x) = - \int f(x) dx$.
The exact solution of Eq.(\ref{eqn1}) is possible only in a few cases: 
more often one only disposes of  approximate results, which usually are obtained by 
applying perturbative methods. Perturbation methods are based on an expansion in 
some small parameter in the problem and the approximate solutions are therefore 
obtained in the form of a polynomial in such a parameter.
This is the case of the Lindstedt-Poincar\'e method and of the multiple-scale 
method, which are widely used in the community.

Unfortunately the validity of the perturbative approaches is restricted to the domain of 
small parameters and the series obtained in such a manner have a finite radius of convergence: 
in other words these series become useless when the parameters are larger than the radius of convergence.
While other techniques are been developed in the literature to deal with this problem, see for 
example \cite{LW02,Mick05}, we wish to present a novel method which was recently devised by one 
of us \cite{Am05a,Am05b,AA06} and which allows one to calculate accurate analytical solutions 
for a classical oscillator, described by Eq.~(\ref{eqn1}). 
The method is based on the powerful ideas of the Linear Delta Expansion (LDE) \cite{lde}, which allows 
to obtain, to any given order,  fully {\sl analytical} and extremely accurate expressions. Since our method
is not based on an expansion in a small parameter, the series obtained converges regardless of the
values taken by the parameter itself.

Since our method is both systematic and analytic it provides several advantages with respect to
other techniques which also claim to work for arbitrary functions $f(x)$: this is the case for example of 
the method of He~\cite{He}, whose shortcomings have been evidentiated by one of us in \cite{San}.

\section{The method}

This problem has been previously studied by one of us in \cite{Am05a,Am05b}, where a nonperturbative
method which allows to calculate the period of the oscillations was devised. 

We will now briefly review the method and then show how to generalize it to calculate the solutions.
For a conservative system the total energy $E$ is constant and the period is simply given by
\begin{equation}
T=\int_{x_{-}}^{x_{+}}\frac{\sqrt{2}}{\sqrt{E-V(x)}}dx,
\label{period}
\end{equation}
where $x_{\pm }$ are the inversion points, obtained by solving the equation $E=V(x_{\pm })$.
Notice that $V(x)=-\int f(x)dx$ is the potential:

As discussed in \cite{Am05a} the problem of calculating this integral can be converted to 
\begin{equation}
T_\delta = \int_{x_-}^{x_+} \frac{\sqrt{2}}{\sqrt{E_0-V_0(x)+\delta  \big[ E -E_0 - V(x)+V_0(x)\big]}}  dx  ,
\label{period1}
\end{equation}
where $V_0(x)$ is a potential chosen to interpolate the original one. 
It is expected that $V_0(x)$ is simpler than $V(x)$.
For $\delta=1$ Eq.~(\ref{period1}) reduces to Eq.~(\ref{period}), whereas 
for $\delta = 0$ this formula yields the period of oscillation in the potential $V_0(x)$. 
We will treat the term proportional to $\delta$ as a perturbation and expand in powers of $\delta$. 
Since $V_0(x)$ depends upon one or more arbitrary parameters 
(which we will indicate with $\lambda$) a residual dependence upon these parameters 
shows up in the period when the expansion is carried out to a finite order. 
In order to eliminate such unnatural dependence we impose the Principle of 
Minimal Sensitivity (PMS)~\cite{Ste81} by requiring that
\begin{equation}
\frac{\partial T}{\partial \lambda} = 0 . 
\end{equation}

We now proceed to define
\begin{eqnarray}
\Delta(x) = \frac{E -E_0 - V(x) + V_0(x)}{E_0-V_0(x)}.
\end{eqnarray}
and write $T_\delta$  as
\begin{equation}
T_\delta = \sum_{n=0}^\infty \frac{(2 n-1)!!}{n!  \ 2^n} (-1)^n  \delta^n 
\int_{x_-}^{x_+} \frac{\sqrt{2} \ \big(\Delta(x)\big)^n}{\sqrt{E_0-V_0(x)}} dx \ ,
\label{period2}
\end{equation}
provided that the series converges uniformly, which is the case
if $|\Delta(x)|<1$ for every $x$, $x_- \le x \le x_+$. We can choose $V_0(x)$
so that each integral in (\ref{period2}) can be performed, thus obtaining an equivalent series 
representation for the original integral (\ref{period}).

Amore and collaborators have applied this method to obtain very precise analytical approximations 
for the period of several oscillator, showing that the error decreases exponentially with the order 
of the approximation \footnote{Recently, Amore and Arceo have given in \cite{AA06} a precise estimate for the rate of 
convergence of the series (\ref{period2}) and they have shown that it converges in all the physical 
region.}.

In particular they considered the Duffing oscillator, which corresponds to the potential 
$V(x) = \frac{1}{2} \ x^2 + \frac{\mu}{4} \ x^4$. The interpolating potential was chosen to be 
$V_0(x) = \frac{1+\lambda^2}{2} x^2$, where $\lambda$ is an arbitrary parameter.
Working to first order they obtained the optimal value $\lambda_{PMS} = \sqrt{3\mu}A/2$ and 
the simple formula:
\begin{equation}
T_{PMS} = \frac{4  \pi}{\sqrt{4 + 3  \mu  A^2}} ,
\label{duff5}
\end{equation}
which provides an error less than $2.2  \%$ to the exact period {\sl for any value of $\mu$ and $A$}.

Physically the PMS allows us to obtain the best potential around which to perform our 
expansion: the form of the potential depends upon the arbitrary parameter $\lambda$ and 
will in general depend on the order to which the calculation is made.

We will now generalize the results of \cite{Am05a,Am05b} by using the conservation of the energy to
obtain the solution to Eq.(\ref{eqn1}). It is straighforward to see that the time spent by the body to
go from $x_-$ to a point $X$ ($x_- \leq X\leq x_+$) is given by 
\begin{equation}
t=\int_{x_{-}}^{X}\frac{\sqrt{2}}{\sqrt{E-V(x)}}dx \ .
\label{X1}
\end{equation}
After repeating the procedure previously explained we have
\beq
t = \sum_{n=0}^\infty \frac{(2 n-1)!!}{n!  \ 2^n} (-1)^n  \delta^n 
\int_{x_-}^{X} \frac{\sqrt{2} \ \big(\Delta(x)\big)^n}{\sqrt{E_0-V_0(x)}} dx \ .
\label{X2}
\eeq
The optimal parameter $\lambda$ will again be chosen according to the same criteria previously adopted.
Notice that, once the integrals in (\ref{X2}) are calculated, one obtains a convergent series representation
for the time $t$ as a function of the position $X$: having proved in \cite{Am05a} the convergence of (\ref{X2}) for 
$X=x_+$, the convergence for $X < x_+$ follows.

As an application we consider the Duffing oscillator and working to first order we obtain
\beq
t=\frac{\left(6 \mu  A^2+8\right) \cos ^{-1}\left(\frac{X}{A}\right)-A \mu  X \
\sqrt{1-\frac{X^2}{A^2}}}{\left(3 \mu  A^2+4\right)^{3/2}} 
\label{xt}
\eeq
or equivalently 
\beq
\Omega_{PMS}\ t = \cos ^{-1}\left(\frac{X}{A}\right)-\frac{A \mu  X \
\sqrt{1-\frac{X^2}{A^2}}}{6 \mu  A^2+8} \ .
\label{Wpms}
\eeq

\begin{figure}
\begin{center}
\includegraphics[width=12cm]{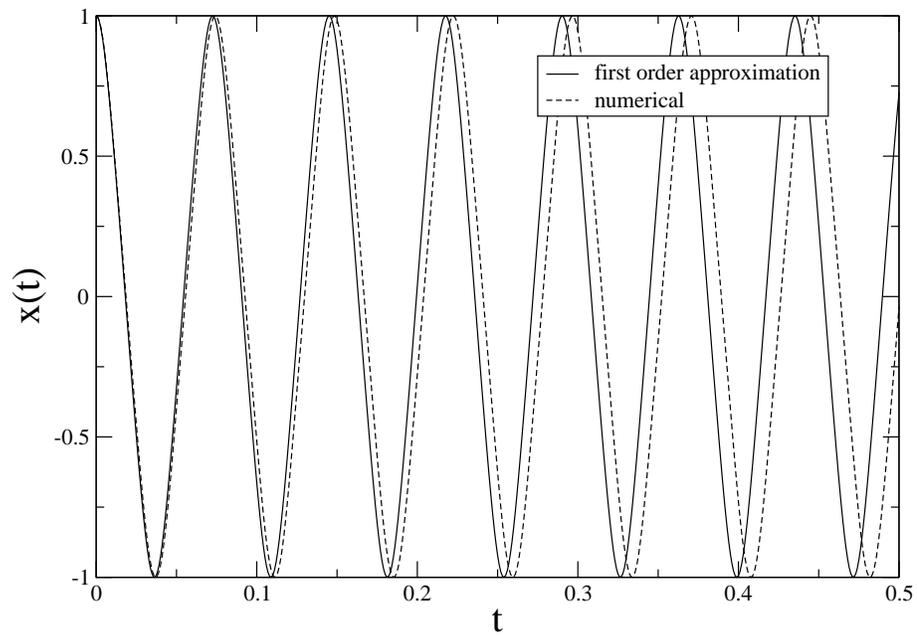}
\caption{Solution corresponding to Eq.~(\ref{xt}) for $\mu A^2 =10^4$ (solid curve).
The dashed curve is numerical.}
\label{Fig_1}
\end{center}
\end{figure}

In Fig.~\ref{Fig_1} we have compared the numerical solution corresponding to $\mu A^2 = 10^4$ with the 
approximation given by Eq.~(\ref{Wpms}): our simple analytical formula provides an excellent approximation,
even in presence of a huge nonlinearity. This regime is clearly outside the region of applicability of perturbative
methods, such as the Lindstedt-Poincar\'e (LP) method.

To ease the comparison with the LP method we notice that the solution to the nonlinear equation will have the general form
\beq
X(t) = \sum_{n=0}^\infty c_n \ \cos \left[(2 n+1) \Omega t\right] \ .
\eeq
We can obtain the Fourier coefficients $c_n$ simply by using Eq.~(\ref{Wpms}):
\beq
c_n &=& -\frac{2}{\pi} \int_{-A}^{A} X \ \frac{dt}{dX} \ \cos\left[(2 n+1)\Omega_{PMS}t\right] \ dX \nonumber \\
&=& -\frac{2}{\pi} \ 
\int_{-A}^{A} X \ \left[ \frac{-7 \mu  A^2+2 \mu  X^2-8}{2 A \left(3 \mu  A^2+4\right) \
\sqrt{1-\frac{X^2}{A^2}}} \right] \nonumber \\
&\cdot&   \cos \left\{(2n+1)\left[\cos ^{-1}\left(\frac{X}{A}\right)-\frac{A \mu  X \
\sqrt{1-\frac{X^2}{A^2}}}{6 \mu  A^2+8}\right]\right\} \ dX \ .
\label{Cn}
\eeq

Eq.~(\ref{Cn}) cannot be evaluated analytically in its present form. However we notice that the function
\beq
\xi \equiv -\frac{A \mu  X \ \sqrt{1-\frac{X^2}{A^2}}}{6 \mu  A^2+8}
\eeq
fullfills the constraint $|\xi | \leq -1/12$ and therefore can be used as an expansion parameter.

Working to order $\xi^3$ we obtain
\beq
\label{fr0}
c_0^{(PMS)} &=& \frac{A \left(26449 \mu^3 A^6+107456 \mu^2 A^4+145408 \mu A^2+65536\right)}{1024 \left(3 \mu  A^2+4\right)^3}\\
\label{fr1}
c_1^{(PMS)} &=&  \frac{A^3 \mu  \left(6435 \mu ^3 A^6+26424 \mu ^2 A^4+36096 \mu  \
A^2+16384\right)}{2048 \left(3 \mu  A^2+4\right)^4} \\
\label{fr2}
c_2^{(PMS)} &=& \frac{5 A^5 \mu ^2 \left(427 \mu ^2 A^4+1112 \mu  A^2+768\right)}{6144 \ \left(3 \mu  A^2+4\right)^4} \\
\label{fr3}
c_3^{(PMS)} &=& \frac{49 A^7 \mu ^3 \left(5 \mu  A^2+16\right)}{12288 \left(3 \mu  \ A^2+4\right)^4}\\
&\dots& \nonumber
\eeq

Notice that the coefficients above are rational functions of $\mu A^2$ and all tend to finite values for $\mu A^2 \rightarrow \infty$.
For $\mu A^2 \ll 1$ the perturbative expressions are found 
\beq
c_0^{(PMS)} &\approx& A \left[ 1- \frac{1}{32} (\mu A^2) + \frac{23}{1024}(\mu A^2)^2 - \frac{1055}{65536} (\mu A^2)^3 +\dots \right] \\
c_1^{(PMS)} &\approx& A \left[ \frac{\mu A^2}{32} - \frac{51 \mu^2 A^4}{2048}+\frac{1287 \mu^3 A^6}{65536}  +\dots \right] \\
c_2^{(PMS)} &\approx&  A \left[ \frac{5}{2048}(\mu A^2)^2 - \frac{745}{196608} (\mu A^2)^3 +\dots \right] \\
c_3^{(PMS)} &\approx&  A \left[ \frac{49}{196608} (\mu A^2)^3 +\dots \right] \\
&\dots& \nonumber
\eeq
which in part reproduce the results obtained with the LP method:
\beq
X^{(LP)}(t) &=& A \left(1 -  \frac{\mu A^2}{32} +  \frac{23 \mu^2 A^4}{1024}  -\frac{547 \mu^3 A^6}{32768} +\dots \right)
\cos\left[\Omega_{LP}\ t\right] \nonumber \\
&+& A \left(\frac{\mu A^2}{32} - \frac{3 \mu^2 A^4}{128}+\frac{29 \mu^3 A^6}{16384} 
+\dots \right)\cos\left[3\Omega_{LP}\ t\right] \nonumber \\
&+&  A \left( \frac{\mu^2 A^4}{1024} - \frac{3 \mu^3 A^6}{2048} + \dots \right) \ \cos\left[5\Omega_{LP}\ t\right] \nonumber \\
&+& A \left( \frac{\mu^3 A^6}{32768} +\dots \right) \  \cos\left[7\Omega_{LP}\ t\right] + \dots 
\eeq

It is useful to consider the exact solution to the Duffing equation which, given the initial conditions used can be cast
in the form
\beq
X^{(exact)}(t) &=& A \ {\rm cn}\left(\sqrt{1+\mu  A^2}\ t | \frac{A^2 \mu }{2 \left(\mu  A^2+1\right)}\right) \ .
\eeq

We can use Eq.(16.23.2) of \cite{AS} to write
\beq
{\rm cn}\left(u|m\right) &=& \frac{2\pi}{\sqrt{m} K(m)} \ \sum_{n=0}^\infty \frac{q^{n+1/2}}{1+q^{2n+1}} \ \cos (2n+1)v
\label{cn}
\eeq
where $q\equiv e^{-\pi K(1-m)/K(m)}$ and $v\equiv \pi u/2K(m)$.

\begin{figure}
\begin{center}
\includegraphics[width=12cm]{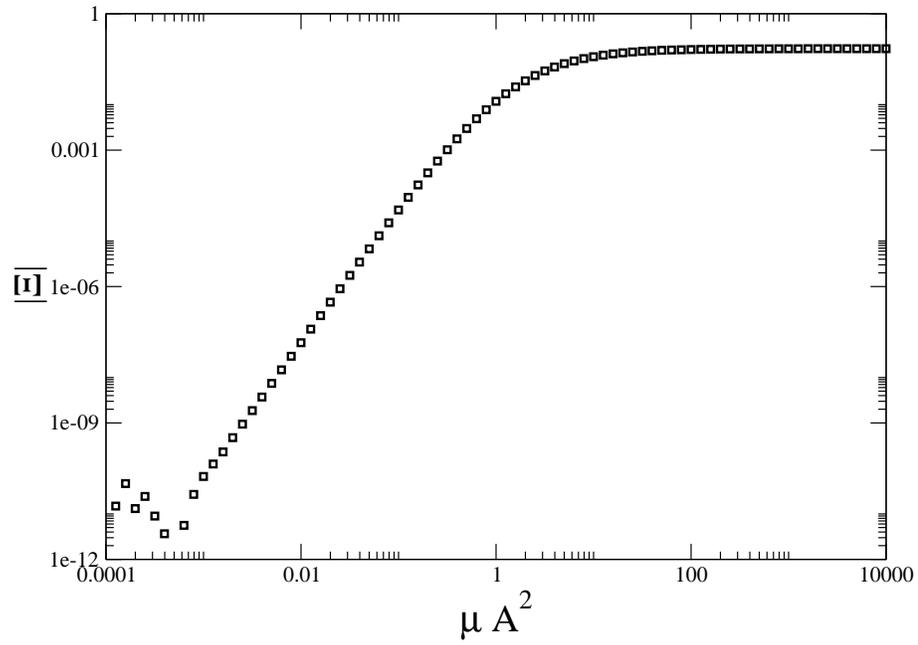}
\caption{Absolute value of the error over the Fourier coefficient $c_0$ using the approximate expression
(\ref{fr0}).}
\label{Fig_2}
\end{center}
\end{figure}

We can easily read off Eq.~(\ref{cn}) the Fourier coefficients of the ${\rm cn}$ function and compare them with the approximations
Eqs.(\ref{fr0}),(\ref{fr1}),(\ref{fr2}) and (\ref{fr3}): the leading Fourier coefficient is reproduced with a
maximum error of $0.15 \%$. In Fig.~\ref{Fig_2} we plot the absolut value of the 
error $\Xi = \left(\frac{c_0^{approx}-c_0^{exact}}{c_0^{exact}}\right) \ \times 100$. We notice that for $\mu A^2 < 1$ the error
is quite small and dies exponentially fast as the limit $\mu A^2 \rightarrow 0$ is approached; on the other hand, in the limit
$\mu A^2 \rightarrow \infty$ the error reaches a plateau. The plateau exists because our approximate coefficient 
$c_0^{approx}$ has the correct asymptotic behaviour, i.e. 
\beq
\lim_{\mu A^2 \rightarrow \infty}  \frac{c_0^{approx}}{c_0^{exact}} =
\frac{26449 e^{-\pi /2} \left(1+e^{\pi }\right) \sqrt{\frac{\pi }{2}}}{110592 \ \Gamma \left(\frac{3}{4}\right)^2} \approx 
1.0017 \ .
\eeq

Although the remaining coefficients are reproduced with less accuracy (for the coefficient $c_1$ we have
a maximum $10\%$ of error) the overall solution is very accurate, since $c_0$ is much larger than all the remaining coefficients.
To prove this statement we can read off the exact expressions for the $c_n^{exact}$ from Eq.~(\ref{cn}) and calculate 
\beq
R_n = \lim_{\mu A^2 \rightarrow \infty} \frac{c_n^{exact}}{c_0^{exact}} = \frac{e^{n \pi } \left(1+e^{\pi }\right)}{1+e^{2 \pi  n+\pi }} \ ,
\eeq
which decays exponentially for large $n$. In the case $n=1$, we see that, even in the asymptotic limit $\mu A^2 \rightarrow \infty$, 
the coefficient $c_0^{exact}$ is about $22$ times larger than $c_1^{exact}$, which is the key to the precision of our results.

Notice that the limit $\mu A^2 \rightarrow \infty$ corresponds to considering a purely anharmonic oscillator 
$V(x) =  \frac{\mu}{4} \ x^4$: clearly, our method is capable to deal quite efficiently also with this case.

It is worth stressing that the present analysis has been carried out only to first order $\delta$: however, since the method is 
geometrically convergent ( see \cite{AA06}) one expects that much higher
precision can be obtained by applying it to higher orders, although this issue is not pursued in this paper. 
We also stress that our method is completely general and that it can be applyied to a large class of potentials:
a detailed analysis of the application of the method to calculate the period of general oscillators is given in \cite{Am05b}.

\section{Conclusions}

We have presented a method to obtain arbitrarily accurate solution for a conservative oscillator.
The particular technique that we used worked for small and large nonlinearities of the equations,
and provides a simple but very accurate approximation. A comparison with the exact solution and with
the perturbative solution obtained using the Lindstedt-Poincare is done. 
It was shown that errors as small as $0.15\%$ are recorded on the leading Fourier coefficient.

The method provides several advantages over other well established methods in the literature: first of all, 
to the best of our knowledge our method is the only nonperturbative method which allows to obtain 
fully analytical results and for which the exponential convergence of the series is proved (\cite{Am05a,Am05b,AA06}); 
secondly, previous work done in \cite{Am05a,Am05b,AA06}, where the method was used to calculate the period (and not the
solution) has shown that our method can be used for a quite large of class of potentials, even in cases where the exact 
result cannot be obtained; finally it is easy to calculate higher order contributions ({\sl never involving special functions})
with our method.

\section{Acknowledgement}
This work has been partially supported by CONACYT, grant 40633.

\newpage
~

\newpage
~

\end{document}